# 3D optical manipulation of a single electron spin

Michael Geiselmann [1], Mathieu Juan [1], Jan Renger [1], Jana M. Say [2], Louise J. Brown [2], F. Javier García de Abajo [3], Frank Koppens [1] and Romain Quidant [1, 4, *]

[1] ICFO - Institut de Ciencies Fotoniques, Mediterranean Technology Park, 08860 Castelldefels (Barcelona), Spain

[2] Department of Chemistry & Biomolecular Sciences, Macquarie University, Sydney, New South Wales 2109, Australia

[3] IQFR - CSIC, Serrano 119, 28006 Madrid, Spain

[4] ICREA - Institució Catalana de Recerca i Estudis Avançats, Barcelona, Spain

* romain.quidant@icfo.es

## Abstract

**Nitrogen vacancy (NV) centers in diamond are promising elemental blocks for quantum optics [1, 2], spin-based quantum information processing [3, 4], and high-resolution sensing [5-13]. Yet, fully exploiting these capabilities of single NV centers requires strategies to accurately manipulate them. Here, we use optical tweezers as a tool to achieve deterministic trapping and 3D spatial manipulation of individual nano-diamonds hosting a single NV spin. Remarkably, we find the NV axis is nearly fixed inside the trap and can be controlled in-situ, by adjusting the polarization of the trapping light. By combining this unique spatial and angular control with coherent manipulation of the NV spin and fluorescent lifetime measurements near an integrated photonic system, we prove optically trapped NV center as a novel route for both 3D vectorial magnetometry and sensing of the local density of optical states.**

While NV centers arise randomly from the presence of natural nitrogen impurities in diamond, they can be artificially created at predefined positions through ion implantation on ultra-pure diamond bulk crystals [13, 14]. Alternatively, single NV centers can also be found in nanocrystal diamonds (ND) whose sizes range from ~5nm to 100 nm [15]. Their nanosize is advantageous because they can be transported to a location of interest, allowing one to place a single spin in very close proximity to photonic [16, 17] and plasmonic [18-23] systems to produce more efficient single-photon sources [19, 23]. Furthermore, a mobile single NV center can act as a nanoprobe of the magnetic field [5-10], the electric field [11] or the electromagnetic Local Density of Optical States (LDOS) [19, 24]. Manipulation of single NV centers has so far been attained by using scanning probe based approaches such as Atomic Force Microscopy (AFM) [5-7, 9, 10, 21, 25]. Although these approaches have led to a remarkable level of control, the experimental implementation is limited to quasi-2D manipulation and requires nanoscale positioning of the ND on the AFM tip. Also, the presence of the host tip can affect the intrinsic properties of the NV center as well as the local fields being probed. In addition, this AFM-based approach typically operates in gas environments, while liquid-phase environments are generally required for bio-applications. Importantly, no technique has so far demonstrated in-situ control over the orientation of the NV axis, which is important for full vectorial magnetic and optical sensing, based on a single spin or optical dipole, and for the implementation of quantum information processing schemes.

In this work, we experimentally demonstrate an novel approach based on 3D optical manipulation of an individual ND hosting a single NV center. We actively select a single ND that is diffusing in a liquid, and once trapped, a high level of control of the spin position of the NV center is achieved along the three spatial axes with nanoscale resolution. Remarkably, the orientation of the trapped NV does not fluctuate over time and can be accurately controlled, which allows us to implement coherent manipulation of a single NV spin and vectorial magnetometry. Finally, we show that lifetime mapping while raster scanning the trapped NV above a patterned surface provides direct insights into the electromagnetic LDOS near an integrated channel waveguide.

The NDs used in our experiments have diameters of about 60-70 nm and are dispersed in an aqueous solution confined in a static fluidic chamber. The optical trap is formed by tightly focusing to the same point two equally intense and collinearly polarized counter-propagating beams from a Nd:YAG (1064 nm) laser (Fig. 1). An enhancement in the stability of the trap results from both the cancellation of the scattering forces combined with the constructive interference along the optical axis. This strategy enables us to trap and manipulate a single ND (n=2.391) as small as 50 nm, in all three spatial directions. In our experiment, optical manipulation of the ND is combined with a full characterization of its NV center. At each spatial position (see **Methods**), we perform time-correlated single-photon counting, extract fluorescence lifetimes and perform optically-detected electron spin resonance (ESR) spectroscopy [26, 27].

In order to deterministically trap an individual ND containing a single NV center, we use a highly sensitive Electron Multiplying Charged Coupled Device (EM-CCD) camera to follow the diffusion of the NDs in real-time (see Figs. 1d, 1e and 2). We immobilize the ND of interest by bringing the focus of the dual-beam trap close to it. To verify that this ND only hosts a single NV center, we perform coincidence measurements using a Hanbury-Brown and Twiss (HBT) detector. Fig. 1a shows measured antibunching below the threshold of 0.5, proving that the trapped ND contains a single NV center. In order to assess the accuracy of our manipulation approach, we resolve the degree of spatial confinement of the ND within the trap (see

**Methods**) and find a full width at half maximum (FWHM) of the Gaussian distribution to be 80nm along both transverse directions (insets of Fig. 2). In the Z-direction (along the beams) we expect its confinement to be smaller than 80nm, due to the interference fringes of the two trapping beams. The achieved trap stiffness is such that the ND can be moved in the chamber over distances as large as 100 µm.

This demonstration of single NV-center trapping grants us access to performing vectorial magnetometry based on a single electron spin. A crucial requirement for this is the stability of the NV-axis (i.e., the spin quantization axis) inside the trap, which we assess through optically-detected ESR measurements (Fig. 3). To this end, we position the ND in the vicinity of a gold wire microwave antenna integrated into the chamber and connected to a microwave generator. The ESR spectrum is obtained by monitoring the NV fluorescence while scanning the microwave frequency over the $m_s$=0 ↔ $m_s$= ±1 2.87 GHz transition. Without any external magnetic field, the $m_s$=±1 states are degenerate, leading to a single dip in the ESR spectrum, as shown in Fig. 3a. In contrast, when we apply an external magnetic field, its vector projection along the intrinsic magnetic dipole of the NV induces Zeeman splitting between the $m_s$= ±1 states. Consequently, the orientation and stability of the NV axis can be determined by monitoring the Zeeman splitting induced by a fixed-amplitude (1.5 mT) external magnetic field along each of the three spatial directions (X-axis shown in Fig. 3b). Fig. 3c shows a Zeeman splitting that is maximum for a specific direction of the field (for this trapped NV: θ=61° and φ=77° in polar coordinates), which corresponds to the NV spin axis.

Long-term stability is evaluated by repeating these measurements after 30 minutes, and we find that the NV orientation is the same within an error range of 10°. This demonstrates that the ND has not substantially rotated within the trap. We attribute this behavior, which was observed for the majority of the trapped NDs (around 80%), to the asymmetrical shape of the nanocrystals (that translates into an asymmetrical polarizability), which creates a preferential equilibrium axis within the linearly polarized trap (Fig. S6 of **Supplementary information**). Consequently, changing the orientation of the polarization axis of the trapping laser should allow us to control the orientation of the NV axis with respect to the optical axis. This additional degree of control over the trapped NV is demonstrated in Fig. 3.d-f, in which we monitor the orientation of the NV axis for three different directions of the trap polarization (for a fixed orientation of the B field). An angular change Δφ in the direction of the field polarization leads to a rotation of the NV axis (projected in the XY plane) by the same angle, therefore permitting us to align the spin orientation along the B field lines to be probed. Thus, this novel manipulation technique is an ideal platform for probing vectorial magnetic fields in three dimensions with nanoscale resolution and a magnetic sensitivity of $40\mu T(\sqrt{Hz})^{-1}$ [8, 29] (see **Supplementary information**).

The fixation of the NV axis, which is associated with the optical dipole axis, also provides an avenue for non-invasive probing of the LDOS. We demonstrate this capability by spatially mapping the NV fluorescence lifetime, which is inversely proportional to the LDOS (projected along the NV axis), in the proximity to a photonic system. For this purpose, a trapped single NV center is raster scanned across a $TiO_2$ channel waveguide patterned on a glass substrate. At each point of the scan (scan step = 15 nm), we measure the fluorescence lifetime of the NV center (Fig. 4a). The fluorescence intensity profile features a maximum above the waveguide. This maximum is attributed predominantly to the significant auto-fluorescence from $TiO_2$ (as can be seen in the EM-CCD fluorescent image Fig. 1d,e)). Conversely, information about the NV-waveguide coupling and the LDOS is provided by the lifetime data in which the NV fluorescence can be easily isolated by applying a proper jitter to the detection. A clear lifetime decrease is observed when the NV is in close proximity to the $TiO_2$ waveguide, which is expected to be inversely proportional to the number of available photon states that couple to the emitting electric dipole of the NV. This quantity is also known as the LDOS, which measures the sum of field intensities of all normalized photon modes as a function of both photon energy and space position. A decrease in lifetime of the NV center is thus associated with an increase in the LDOS. In Fig. 4b, we show the spatial dependence of two components of the LDOS: radiative emission far from the system (upper curve) and non-radiative emission into a guided mode of the waveguide (lower curve). While the former seems to depend only mildly on position, the waveguide mode contribution shows a peak that follows the observed decrease in lifetime. As shown by the red line in Fig. 4a, the theory is in very good agreement with the experiment.

Spatial and angular manipulation of optically trapped single NV centers is a simple and powerful approach that could apply/benefit to a wide range of research. Beyond opening new opportunities for nanoscale magnetic and optical sensing in three dimensions and in liquid environments, it is non-invasive and is potentially applicable to cell biology, enabling imaging of single nuclear spins inside living cells [30, 31]. We also foresee that an extension to multiple traps can be used for the implementation of flying qubits [32] and nanoscale control of spin-spin interactions, materializing a significant advance in scalable quantum computing and simulation schemes.

During submission of our manuscript we came across a work by Horowitz et al. on ESR resonance of multiple optical trapped nanodiamonds with ensembles of NV centers [33].

**Acknowledgements**

This work was partially supported by the Spanish Ministry of Sciences under grants FIS2010–14834 and CSD2007–046-NanoLight.es, the European Community's Seventh Framework Program under grant ERC-Plasmolight (259196), and Fundació privada CELLEX. We thank Dr. Torsten Gaebel and Dr. Andrews Edmonds for fruitful discussions.

**Methods**

**Optical trapping and fluorescence read out.** A near IR ($\lambda$=1064 nm) 1.5 W laser is split by a 50:50 beamsplitter into two beams and focused through two high NA water immersion objectives to a single point. The power at the objective entrance is fixed to 150 mW in each beam to compensate for the objective low transmission at this wavelength (30%).
The NDs were diluted in a 5:1 glycerol:water solution. A droplet of this solution between two coverslips served as a liquid chamber fixed on a XYZ-piezo stage. The glycerol was chosen to increase the viscosity of the medium and consequently lower the Brownian motion of the NDs. An EM-CCD camera was used to detect the fluorescence of a single NV center, and then track single NV centers, trap them and position or scan them at desired locations in the chamber. A 532 nm laser was used to excite the NV center. The emitted fluorescence was detected, after a dichroic filter and a long pass filter at 632 nm, with two APDs.

**Trap characterization.** Using standard position measurements based on a Quadrant Photo-Detector (QPD) [28], we are able to retrieve the position histogram along each of the two directions X and Y perpendicular to the optical axis. The full width at half maximum (FWHM) of the Gaussian distribution is of the order of 80 nm along both directions (Fig. [2]). Since position detection is known to lack accuracy for determining the trap confinement in the Z direction, we can only provide an estimate based on the spatial confinement of the interference fringes. Assuming that the ND remains localized in one fringe at the focus, we expect its confinement to be smaller than 80 nm.

**Nanodiamond preparation.** The nanodiamond sample (Microdiamant, MSY 0-0.1 µm) was refluxed for 6 days at 90°C in a 9:1 ratio of concentrated sulphuric to nitric acids. The nanodiamonds were collected by centrifugation and ultrasonicated using a tip probe sonotrode (Branson, 101-063-346) for 1 hour before being additionally refluxed under the previous conditions. The nanodiamonds were then washed with deionised water and ultrasonicated in 0.1M sodium hydroxide for 1 hour, rinsed with water and re-ultrasonicated in 0.1M hydrochloric acid. The sample was rinsed and resuspended in deionised water. Finally, several centrifugation steps were used to size select the required nanodiamond particles and Dynamic Light Scattering (Malvern Instruments Zetasizer NS) measurements were used to determine the size distribution of the particle to be 74±15 nm.

# FIGURES

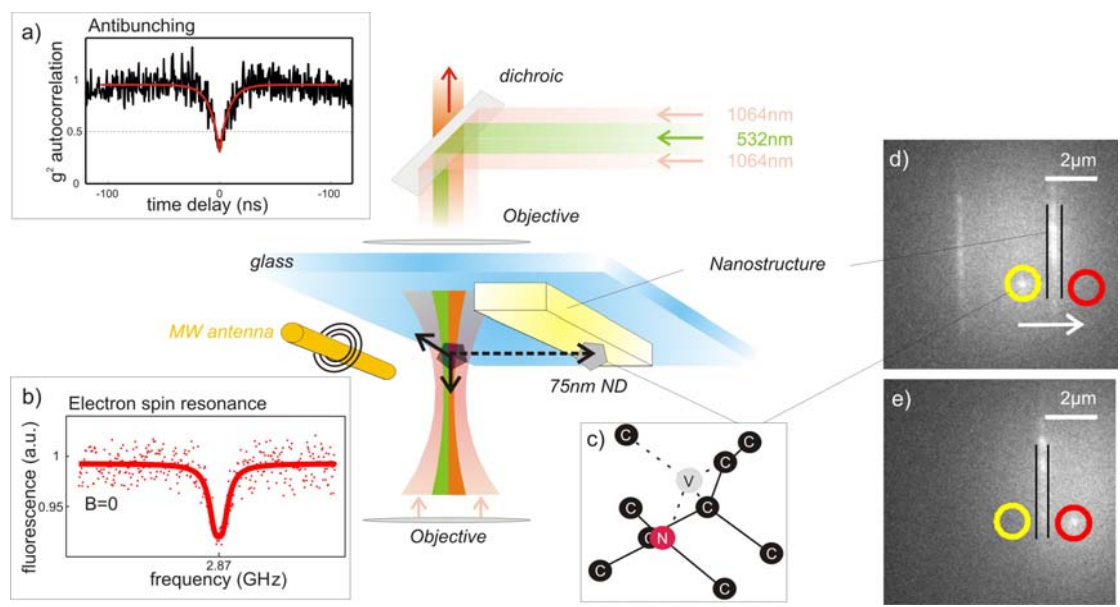

Fig 1. **3D Optical trapping and manipulation of a single electron spin**. A near-IR laser (λ=1064 nm, shown in light red in the central sketch) is focused through two high-NA objectives to trap a nanodiamond (ND) containing a single NV center. A superimposed 532 nm laser (shown in green) excites the NV. The emitted photoluminescence (orange) passes a dichroic filter and is detected with two Avalanche Photo Diodes (APDs). The central schematic shows the basic elements of our experiment. a) Coincidence measurements of a trapped ND confirming the single-photon character of the NV luminescence ($g^{(2)}(0)<1/2$). The red line is an exponential fit with a lifetime decay of 18 ns. b) Optically detected magnetic resonance measurement of a single NV center. c) Schematic of the nitrogen-vacancy (NV) center in diamond consisting of a substitutional nitrogen atom (N) adjacent to a vacant lattice site (V) within the carbon (C) lattice. d), e) Positioning of a single NV center relative to a 250 nm thick $TiO_2$ waveguide. The image is taken with an EM-CCD camera while illuminating a large area with a 532 nm laser. The bright spot is the photoluminescence from the NV center, while the bright lines correspond to fluorescence from the $TiO_2$ waveguide.

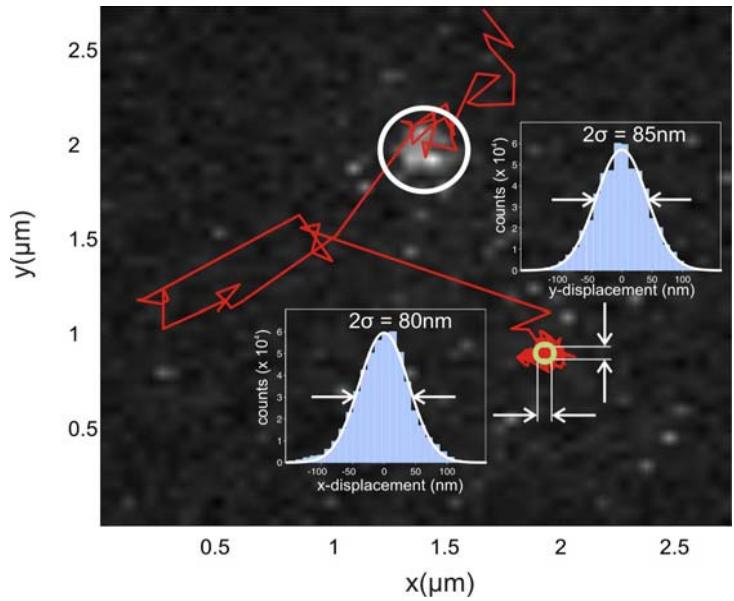

Fig 2. **Tracking and trapping of a single ND with a single NV-center.** EM-CCD image of a fluorescent nanodiamond (white circle) and its position trace (red line). Tracking the fluorescence makes it possible to direct the nanodiamond with the XYZ piezoscanner close to the trap, where it is captured and confined. The insets show position histograms of the time trace of the trapped ND retrieved from position tracking with a quadrant photo diode [28]. The solid blue lines are Gaussian fits to the position histograms. The standard deviation is ~40 nm. The particle is confined within less than 80 nm during 70% of the time.

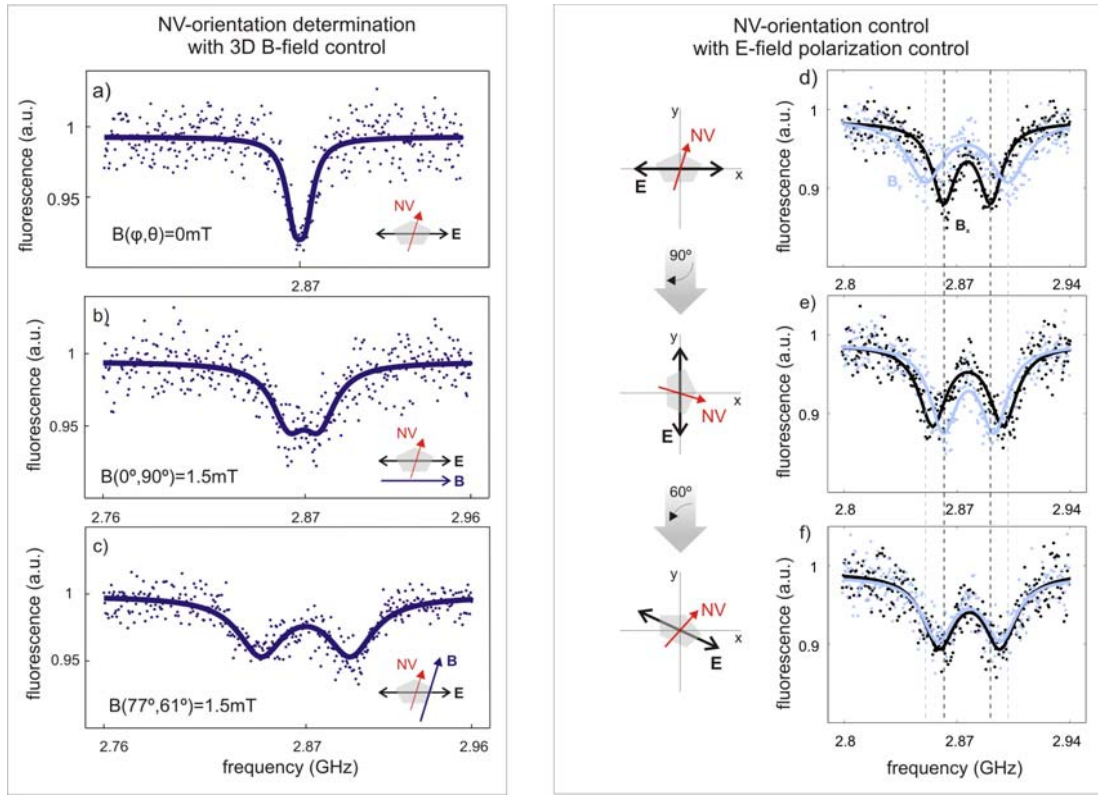

Fig 3. **Vectorial magnetometry and orientation control with an optically trapped single spin.** a)-f) Optically detected electron spin resonance spectra of a trapped ND with a single NV-center. a)-c) Determination of the NV-axis orientation. a) No magnetic field is applied. The $m_s = \pm 1$ ground state is degenerate. b) For the X projection of the magnetic field onto the NV axis only a small Zeeman splitting is visible. c) When the magnetic field is aligned with the NV axis, a maximal splitting is recorded. This makes vector magnetometry possible. d)-f) Control of the NV-axis orientation by rotating the polarization of the trapping laser. A rotation in polarization by an angle Δφ rotates the NV-axis by the same angle in the XY-plane as monitored by the change in $\vec{\mu} \cdot \vec{B}$. The dashed lines indicate the position of the magnetic field splitting in d) as a guide to the eye. (Note: the ground state for this NV was already degenerated for B=0.)

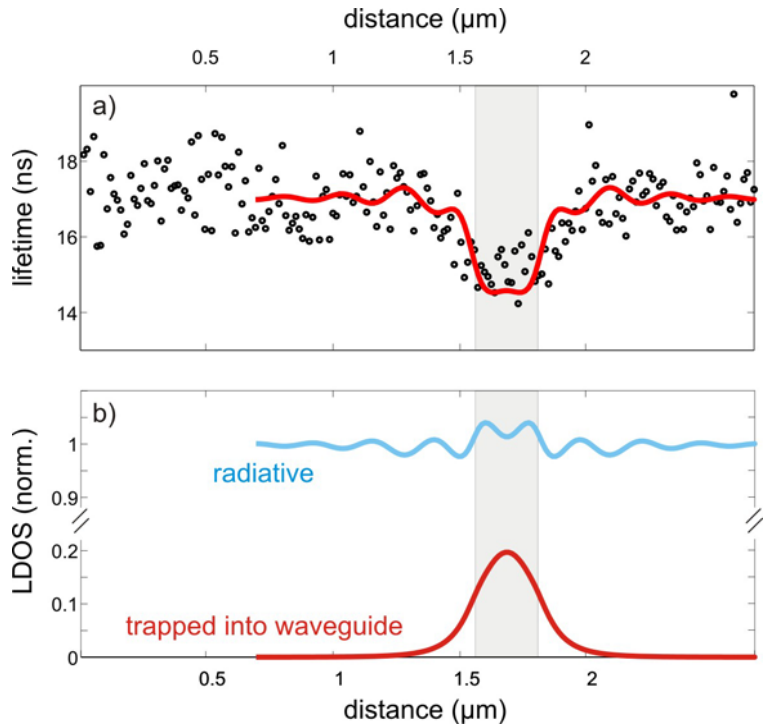

Fig 4. **LDOS mapping across a TiO$_2$ waveguide** a) Fluorescence lifetime of the NV center, which were acquired at different points across and over the waveguide (average over 2 consecutive scans). The red line shows the calculated lifetime of a single emitter scanned 50 nm above a 250 nm TiO$_2$ waveguide. b) Calculated radiative and waveguided contributions of the LDOS for a dipole scanned 50nm above a 250 nm TiO$_2$ waveguide. The radiative part has only a small contribution in comparison to the non-radiative part that is trapped in a waveguide mode and is responsible for the change in lifetime of the NV center. The shaded area corresponds to the position of the waveguide.

## Supplementary Information

**Optically detected ESR and determination of the NV-axis**

In the fluid chamber an implemented gold wire serves as a microwave antenna that is connected to a microwave generator and amplifier. The static magnetic field is created by 3 independent coils with axes along X, Y and Z directions. The coils were calibrated with a Hall probe. The first microwave scan is made without any magnetic field. The microwave generator sweeps 15 times from 2.76 GHz to 2.96 GHz in steps of 0.5 MHz. To compensate fluorescence fluctuations, the APD detects the fluorescence signal with and without microwave irradiation. The two signals are divided and averaged over 20 times. A dip is observed in the fluorescence signal each time the microwave frequency hits the $m_s=0 \leftrightarrow m_s=\pm1$ transition frequency of the ground state. Next, a magnetic field B=1.5 mT is applied along the X direction. The $m_s=\pm1$ state is Zeeman split, but only experiences the projection from the B field onto the NV spin axis. Consecutively, a magnetic field of the same amplitude is applied along the Y and Z directions, respectively. Comparing the relative splitting Δ for fields along each axis, the contribution of the respective magnetic field is determined and we calculate the polar angles corresponding to the orientation of the NV axis as

$$\varphi = \mathrm{atan}(\Delta y/\Delta x) \qquad (1)$$

$$\theta = \mathrm{atan}[\Delta x/(\Delta z \cos\varphi)] \qquad (2)$$

where $\Delta x$, $\Delta y$ and $\Delta z$ are the Zeeman splittings produced by magnetic fields along X, Y and Z directions, respectively. In addition to the vector spin direction, this measurement demonstrates that the particle is not substantially rotating in the trap. Because the measurement is based on averaging, the different projections of the magnetic field on the NV axis would result under rotation in substantial broadening of the ESR resonance, preventing the two different transitions from being resolved.

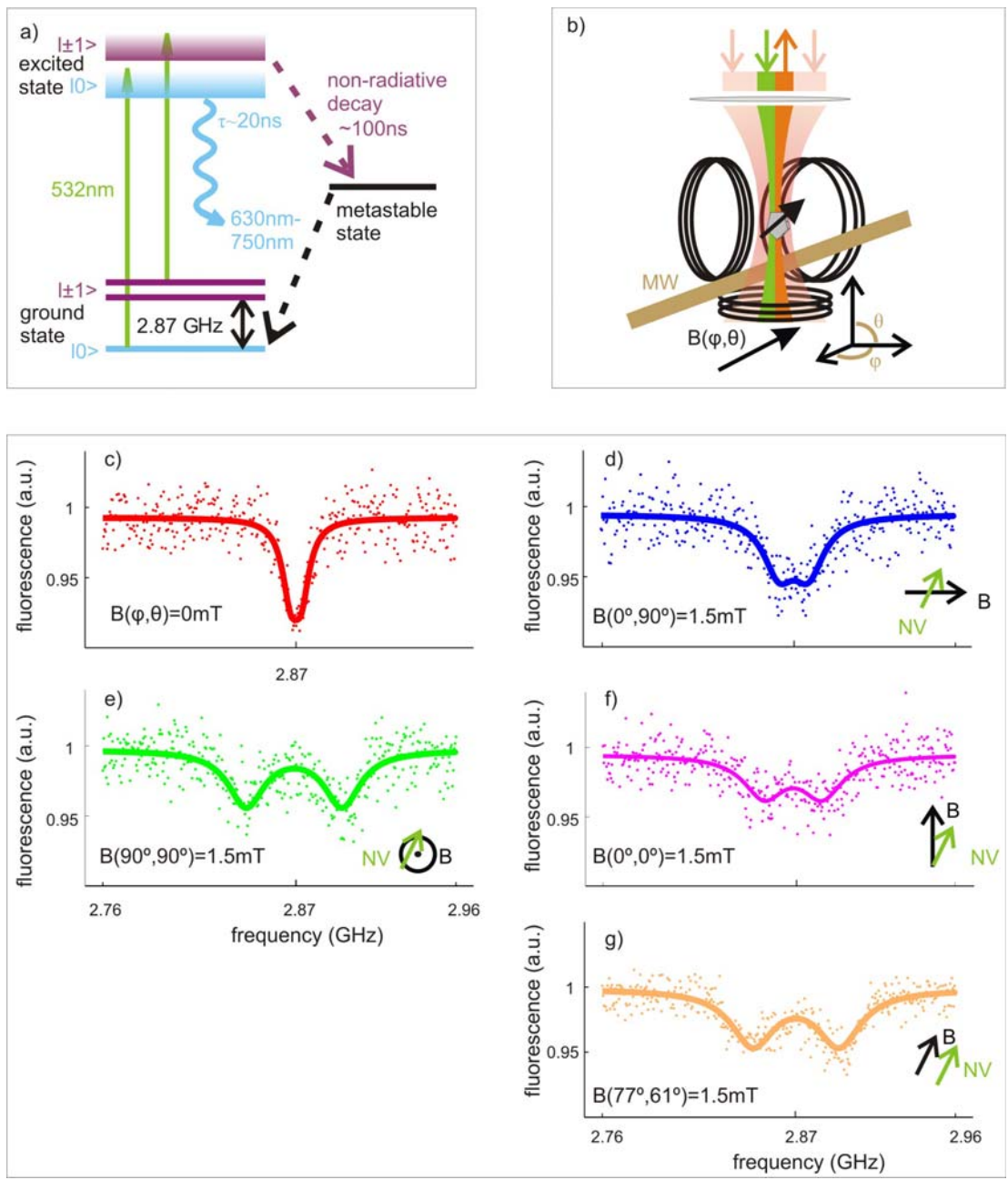

Fig S5. **ESR spectroscopy to determine the orientation of the NV axis.** a) Energy level diagram of a NV center in diamond. Spin-selective optical transitions can be monitored by the fluorescence of the NV center. Non-radiative decay produces a significant reduction in the photoluminescence, resulting in a drop in the fluorescence when a spin transition from $m_s=0 \leftrightarrow m_s=\pm1$ is induced by the microwave field. b) Sketch of the experimental setup: three coils provide a fixed magnetic field that can be orientated in all three spatial directions.

c)-g) Each graph shows a microwave frequency scan across the ground state transition frequency $m_s=0 \leftrightarrow m_s=\pm1$. No magnetic field is applied in c), so that the $m_s=\pm1$ ground state is degenerate. In d), e), f) a magnetic field of 1.5 mT is applied along the X, Y and Z axes, respectively. Different Zeeman splittings are observed, depending on the projection of the external field onto the NV axis. The actual orientation can be retrieved according to Equations (1) and (2). g) In a subsequent ESR scan, we observed a maximal Zeeman splitting by applying a magnetic field of the same amplitude (1.5 mT) along this calculated direction, thus corroborating the validity of our spin-orientation retrieval procedure.

**Magnetic sensitivity**

We estimate the magnetic field sensitivity of our experiment to be

$$\eta = 0.77\, h\, \Delta\nu\, (2\, \mu_B\, C\, \sqrt{R})^{-1} = 40\, \mu T\, (\sqrt{Hz})^{-1},$$

where h is the Planck constant, $\Delta\nu$ is the broadening of the ESR dip, $\mu_B$ the Bohr magneton, C is the contrast for the ESR and R is the count rate [1]. Incidentally, according to Taylor *et al.* [2], magnetic sensitivities for nanodiamonds with a dephasing time $T_2^* = 1$ µs are possible down to 1µT $(\sqrt{Hz})^{-1}$. This can be reached by implementing a Ramsey type magnetometry [2] or by avoiding power broadening [1].

**Scanning Electron Microscopy (SEM) of the nanocrystal diamonds**

The nanodiamonds we use have asymmetric morphologies as shown by SEM on an evaporated droplet of a concentrated ND solution. An example of such images is shown below. The asymmetry is beneficial to assist the stability of the ND within the trap by providing a preferential orientation.

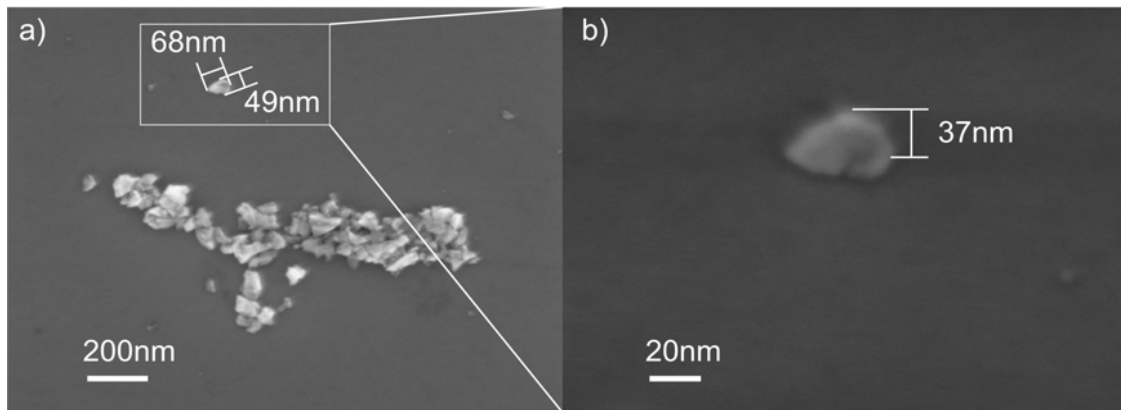

Fig S6. **Morphology of the ND**: a) SEM image obtained on an evaporated droplet of a concentrated ND solution. The size distribution is in agreement with the Dynamic Light Scattering measurement (see Methods section in main paper). Most of the crystals have an asymmetry, which prevents rotation in the optical trap. b) Tilted SEM image of a single ND. This crystal has a size of 68 nm x 49 nm x 37 nm.